# Technology Review of Blockchain Data Privacy Solutions

# Market Research Document

**Authors**
Jack Tanner, Blockchain and SSI developer at Jack and the Blockstalk
Roshaan Khan, Product Manager at Block One

**Contributors**
Alexandre Bourget, Co-founder and CTO at Dfuse
Andres Gomez Ramirez, Security Researcher at EOS Costa Rica
Brendon Ross, Co-founder at Rewired.one
Duncan Westland, Head of Global Blockchain R&D at Ernst & Young
Edgar Fernandez, Co-founder at EOS Costa Rica
Felix Shnir, Quorum Engineering at JPMorgan
Jeremy Nation, Content Writer at Block One
Mark Woods, VP of Product Management at Block One
Max Gravitt, Blockchain Lead and Founder at Digital Scarcity
Michael Harris, Product Manager at Block One
Paul Sitoh, Freelance Consultant at Applied Consentia
Rhett Oudkerk Pool, CEO at Europechain & EOS Amsterdam
Sneha Damle, Developer Evangelist at R3
Suneet Bendre, Software Engineer Tech Lead at RobustWealth
Torben Anderson, Co-founder at Rewired.one
Xavier Fernandez, Blockchain Developer at EOS Costa Rica
Yannick Slenter, Digital Strategist at Europechain & EOS Amsterdam

**Reviewers**
Amanda Clark, Manager of Product Management at Block One
Caspar Roelofs, Founder at Gimly
Daniel Larimer, Former CTO at Block One
Ian Holsman, VP of Software Engineering at Block One

# Abstract

This objective of this report is to review existing enterprise blockchain technologies - EOSIO powered systems, Hyperledger Fabric & Besu, Consensus Quorum, R3 Corda and Ernst & Young's Nightfall - that provide data privacy while leveraging the data integrity benefits of blockchain. By reviewing and comparing how and how well these technologies achieve data privacy, a snapshot is captured of the industry's current best practices and data privacy models. Major enterprise technologies are contrasted in parallel to EOSIO to better understand how EOSIO can evolve to meet the trends seen in enterprise blockchain privacy. This research was done to guide the working group in their decisions around a data privacy architecture for EOSIO and the results are shared publicly for the benefit of all.

Every technology reviewed, besides EOSIO, has a managed privacy offering that is meant to meet requirements of enterprise users. The approach to meeting this privacy standard is varied and unique for each project. The report compared technologies with an analysis framework taking into account maturity, data privacy relationships, private data transport, private data integrity, and data privacy.

The following strategies and trends were generally observed in these technologies:

**Cryptography** - the hashing algorithm was found to be the most used cryptographic primitive in enterprise or changeover privacy solutions.

**Coordination via on-chain contracts** - a common strategy was to use a shared publicly ledger to coordinate data privacy groups and more generally managed identities and access control.

**Transaction and contract code sharing** - there was a variety of different levels of privacy around the business logic (smart contract code) visibility. Some solutions only allowed authorised peers to view code while others made this accessible to everybody that was a member of the shared ledger.

**Data migrations for data privacy applications** - significant challenges exist when using cryptographically stored data in terms of being able to run system upgrades.

**Multiple blockchain ledgers for data privacy** - solutions attempted to create a new private blockchain for every private data relationship which was eventually abandoned in favour of one shared ledger with private data collections/transactions that were anchored to the ledger with a hash in order to improve scaling.

We researched and identified insights based on common problems, solutions and strategies seen in the architecture of the reviewed privacy solutions. The insights will be used to drive the working group's proposal for an EOSIO based enterprise privacy solution. The research overwhelmingly points to there being no end-all, be-all, panacea to solve private data on blockchain networks. Every solution has benefits and drawbacks while different use cases have varying needs. The most complete solution is paralleled most by the Ethereum Enterprise Alliance defining the standards for a "privacy stack" that lays the foundation for developers to implement privacy in a way that benefits them the most.

# Table of Contents

# 1 Introduction

This report is written for technical professionals working or planning on working with blockchain and other high data integrity solutions with data privacy requirements. This report reviews and summarizes trends in data privacy technologies utilizing cryptographically secure blockchain infrastructure.

"Privacy and scalability are the foremost reasons why organizations choose private blockchains over public blockchains," based on research conducted by Hyperledger [1].

Blockchain technology introduces a new architecture for applications that enables greater transparency, efficiency, security, data integrity, and traceability while reducing the need for third parties to oversee transactions. [2] Enterprises have found several use cases for blockchain to improve upon current processes and solve existing problems in new ways.

Well known public blockchains like Bitcoin or Ethereum have data privacy limitations due to their inherently transparent nature. Most public networks, due to their transparent-by-design nature, do not fit business privacy requirements. Fully public environments go against the standard most enterprises focus on protecting their customers', clients', and their own business's information.

Private blockchains remove the public transparency of data and limits access to a set group of approved participants. This approach can work for some use cases but even within an organization's stakeholders, especially across different businesses, there is still a need for privacy and the ability to secure private information through a single organization.

Coming to the question of which blockchain approach is better, a public blockchain seems to stand out as the best option as it can be applied in most use cases and is free from restricted access. [3] In addition, our interviews with enterprises using blockchain has even shown that some enterprises would prefer to use public blockchains for their applications if they could still maintain a level of privacy while doing so. However, private blockchains do serve plenty of use cases and due to BAAS providers like AWS or Azure, they are easy to set up.

While blockchain technology can support both private and public blockchains the market has shown the need for a privacy stack [4] to be able to supplement both and enhance enterprise useability of the technology.

The problem is achieving privacy in transparent systems. There are many enterprise benefits to operating more transparently, but there remains desire for some degree of privacy. Fully transparent systems have many benefits, but some enterprises do not want to, or cannot operate transparently with all their data.

Therefore, in order for blockchain systems to be able to compete with traditional systems there needs to be a degree of data privacy while maintaining the data integrity properties of a blockchain.

User's data privacy should be the default and considered a human right. Exposing a user's privacy should be justified, not the other way around. In the enterprise world there are strict government regulations around a user's private data that must be abided by. Specific use cases involving personal information have driven the creation of Self Sovereign Identity standards and an architecture that leverages a transparent blockchain for decentralized public key infrastructure. There is a strong emphasis on privacy in these new identity standards.

Enterprises are also able to draw competitive advantages to capitalize on their users' data. For blockchain to be taken more seriously as a solution for enterprise use cases, privacy needs to be addressed in a manner that is beneficial to businesses.

## 1.1 EOSIO data privacy working group

The need for a privacy stack has led to the birth of the EOSIO data privacy working group between Block One and major community players in the EOSIO ecosystem to research, propose, and implement a privacy solution for business privacy use cases. As our first step we interviewed enterprises utilizing private data in blockchain and conducted a market review of what major technologies offer their users. This report is the cross-market comparison of major data privacy technology and their data privacy characteristics.

The working group's objective is to establish a perspective on the current state of enterprise privacy tooling in the blockchain industry.

This report's objective is to review existing tech - both EOSIO and non-EOSIO - that provides data privacy while leveraging the data integrity benefits of a blockchain. We believe this information is holistically beneficial to the blockchain industry and contributes comprehensively to the practice of on-chain privacy. Especially for enterprises that want to compare blockchain offerings and understand what kind of data privacy capabilities are offered to them before they commit to building. This report is a cross examination of existing solutions, Hyperledger Fabric, Hyperledger Besu, Quorum, Corda, and EOSIO, as well as the Zero Knowledge Proofs (ZKPs) solution, Nightfall.

This report is being released openly and freely to share insights on the data privacy landscape with the EOSIO and greater blockchain community, because of the group's belief that privacy on blockchains will be a major topic in coming years for businesses interested in utilizing the technology. By providing valuable insights and information about the current technology market and gaps for the technology industry the group hopes to identify opportune enterprises to join the working group's endeavor and aid in setting the standards for blockchain privacy in business.

Lastly, to lay the groundwork for the forthcoming proposal the working group will make to the open source EOSIO community and technology stakeholders to address private data transactions on the blockchain.

We have used publicly available documentation and sometimes reviewed code bases. We had this content externally reviewed by specialists working with the reviewed technology. However, due to the significant complexity of the technologies reviewed we acknowledge there may still be incorrect facts. Please provide any feedback to roshaan.khan@block.one, thank you for reading.

## 1.2 Intro to EOSIO software

EOSIO is a next-generation, open source blockchain protocol with industry-leading transaction speed and general utilization. Introduced in May 2017, it has since been widely recognized as the first performant blockchain platform for businesses across the world. [5]

Built for both public and private use cases, EOSIO is customizable to suit a wide range of business needs across industries with role-based security permissions, industry-leading speeds and secure application processing. Building on EOSIO follows familiar development patterns and programming languages used by existing non-blockchain applications so developers can create a seamless user experience using development tools they already know and love. [5]

**Some important features of EOSIO**
- EOSIO is governance agnostic because it uses asynchronous byzantine fault tolerance (ABFT) [5]. So, while the EOS public blockchain uses delegated proof of stake, any other governance model, such as democratic, or proof of stake could be plugged into a blockchain developed on EOSIO. This is important in data privacy applications to allow implementers to choose the best governance of their data.
- Through EOSIO's unique architecture the technology has been able to significantly improve throughput[1] and low latency (0.5s blocks) for transactions that has led to its reputation as an enterprise grade blockchain. This scaling has been achieved while keeping the same security model of Bitcoin and Ethereum, where every node cryptographically audits every transaction. This scalability is key for

---

[1] Tested to 9,500 transfer transactions/s [85] - note that throughput depends on usage type

- applications that want to get to production and offer sleek user experiences. This is important for data privacy since certain privacy technologies require intense computations, a fast chain helps keep private transactions usable.
- EOSIO continues to upgrade its capabilities consistently enough to keep it as a front running blockchain as exemplified by ranking number on China's Ministry of Industry and Information Technology's (CCID) Global Public Blockchain Technology Assessment Index [6] several publications in a row. The consistent development done by Block One and the supportive community means the technology doesn't fall behind overtime making it an ideal platform to start building for the future on.

### 1.2.1 EOSIO vs EOS

For those new to EOSIO that may have heard of EOS, it is important to caveat that there is a distinct difference between the two [7]. EOS is a public blockchain governed by a public community, while EOSIO is an open source technology maintained by Block One that can be used to develop any type of blockchain. EOSIO is the protocol that powers EOS which has been customized significantly, can support different governance models, and has similar characteristics to other EOSIO blockchains while remaining fully protocol-compatible [5].

### 1.2.2 Success of EOSIO projects

The EOSIO ecosystem is more than just the EOS public network. There are over a dozen public and many more private projects built by utilizing the technology ranging from games and exchanges, to full blown blockchain development systems that leverage EOSIO's core libraries. The following is a taste of some of the projects built using EOSIO and the wide range of consumer products that have been created.

- Searching and Processing Blockchain Data with dfuse [8]
- WORBLI is a Financial Services Oriented Blockchain Built on EOSIO [9]
- GeneOS Taps Blockchain to Unlock Health Data Insights and Incentives [10]
- Sense.Chat Brings Privacy and Value to Messaging on Blockchain [11]
- AdNode Improves Digital Advertising with EOSIO [12]
- The Ultra Blockchain Network is Built for The Video Game Industry [13]
- Upland Blurs the Boundaries of the Real and Virtual Worlds [14]
- Everipedia's Marriage of Brainpower and Blockchain [15]

## 1.3 Understanding the use case behind privacy

In order to understand the topic at hand a framing use case needs to be presented to visualize how this applies to the real world.

The question to understand when reviewing blockchain privacy solutions is "How can we have data privacy while also maintaining the data integrity properties of blockchain technology?"

For our use case we will be using MediLedger [16], an enterprise blockchain consortium improving pharmaceutical tracking with Pfizer, Walmart, McKesson, AmerisourceBergen, and Cardinal Health as members. They conducted an FDA Pilot Project [17] which will be the basis of this case.

### 1.3.1 Summary of the pilot

The FDA is running a pilot program to help the industry move toward meeting the 2023 requirements of the Drug Supply Chain Security Act (DSCSA) which requires the pharmaceutical industry to be able to track legal changes of ownership of pharmaceuticals in the supply chain. The MediLedger Pilot Project, a consortium of leaders from 25 pharmaceutical companies consisting of industry giants such as Pfizer and Walmart.

The consortium wanted to use a blockchain for the immutability and governance benefits in comparison to traditional systems. The pilot needed to be able to prove to government authorities that drugs going through the supply chain were accurately accounted for and there were untampered records for every change of ownership.

The data privacy challenge surrounded maintaining transaction privacy between two companies in a permissioned environment where all 25 companies had access to the main chain, and therefore all occurring transactions.

### 1.3.2 Why a blockchain was used to fulfil FDA requirements

1. A single blockchain matched the performance and scalability of the currently used G1 standard Electronic Product Code Information Services (EPCIS). There is a business requirement to be able to demonstrate 2000 transactions per second.
2. Data privacy requirements were be met using Zero Knowledge Proof technology which allows for nodes to be hosted by several unique parties while maintaining strict transactional privacy and still ensuring immutability of the ledger.
3. The pilot required an open database where several actors could read and write data without always revealing every aspect of their activities to everyone. Ex: Walmart wants to buy medicine from Pfizer but does not want to reveal information to McKesson about the transaction.
4. A blockchain system validated the authenticity of product identifiers and facilitated the provenance of saleable units back to the original manufacturer.
5. The authenticity of the drug transaction information was confirmed with each transaction allowing for expedited suspect investigations and recalls.
6. The group believes that should a blockchain ecosystem be created as a possible solution to the DSCSA interoperable solution requirement, it should have an open system architecture with appropriate governance to oversee the function of the system and ensure compliance with industry agreed business rules and standards of operation.
7. Governance can be defined by the industry itself.
8. The trust established by a blockchain system can be leveraged for a myriad of additional business applications to the pharmaceutical industry, allowing for compounding benefits for this industry once such a platform is established.
9. The long-term success of an interoperable blockchain solution will require strong participation and adoption from all industry stakeholders.

### 1.3.3 Solution overview

The solution used 3 core technologies:
1. **Private messaging** between clients to exchange confidential messages between trading partners by leveraging EPCIS technology and standards.
2. **A blockchain as a shared, immutable ledger** to register the proof of the authenticity of transactions and execute smart contracts. The blockchain enforced business rules, such as permitting only one company to have legal ownership of a serialized unit at a given time (no double transfer).
3. **zk-SNARKs for enhanced privacy** by ensuring no business data is revealed.

### 1.3.4 What we learned from this case study

1. There is a need for privacy in enterprise blockchain use cases, even in the case where the chain is private and not public.
2. There is a demand for a scalable, interoperable solution where privacy is not a hindrance on performance and throughput.
3. A privacy stack is necessary for blockchain technology to be considered as a general enterprise solution.
4. Blockchain technology has the potential to meet certain market requirements better than traditional solutions.
5. Blockchain platforms are acceptable for government entities looking for greater traceability, accountability, and immutability.

## 2  Review of existing solutions

This section will provide a short description of several promising data privacy solutions and some of their important data privacy characteristics.

This report focuses on data privacy and may simplify other concepts for the sake of understanding the data privacy aspects.

| Project | Chain - Privacy Type | Summary | Main strengths | Main weakness |
|---|---|---|---|---|
| Fabric | Private Chains - Private Transaction Manager using "Data Collections" | Private data collections, allow a defined subset of organizations on a ledger the ability to conduct private transactions without having to create a separate ledger. | Data privacy controls can easily be implemented at a ledger or data collection level. Code from contracts on the shared ledger is only visible between participating organizations. Ability to manage complex private data relationships. | All members in a private collection or ledger can see all past, present, and future transactions in that channel. Read/write set and transaction input data are susceptible to dictionary lookups if simple and without salt. |
| Corda | Public Chains - Sharded Ledger with Private UTXO Transactions | Each node can only see the transactions that they are a participant with or have been explicitly shared. They cannot see anything not permissible thus they effectively simulate a private environment. | Transaction data (not even a hash) is not shared with peers that are not part of the operations. | At least one third party notary is needed to guarantee correctness of transactions and gets full visibility over an application's business logic. Transaction participants are able to see information about a transaction's inputs that they may not have been a participant in. |
| Besu | Private Chains - Private Transaction Manager using "Orion" | Ethereum client that has a privacy manager called Orion. It creates private databases handling private transactions and stamping hashes on a shared Ethereum chain. | Compliance with Ethereum enterprise alliance standards. Manages privacy groups automatically using on chain coordination. | Identity and activity of private transactions in Orion can be discerned under certain conditions. |
| Quorum | Public and Private Chains - Private Transaction Manager using "Tessera" | Ethereum client that has a privacy manager called Tessera. It creates private databases handling private transactions and stamping hashes on a shared Quorum chain (A slightly modified Ethereum protocol). | Compliance with Ethereum enterprise alliance standards. Transaction privacy is defined at a transaction level (there are no privacy groups) making it privacy dynamically managed. | Identity and activity of private transactions in Tessera can be discerned under certain conditions. |
| Nightfall | Public/Permissionless - Zero knowledge proofs | Enables zero knowledge proofs to be conducted on the Ethereum public network. | Strong guarantees of user control privacy. | Expensive to run individual transactions so they batch several together. Difficult to upgrade without all users involved. Experimental tech that becomes outdated quickly. |

Table 1: Summary table of reviewed existing solutions

## 2.1 Common language

**Node** - The computer that runs the data privacy software. This can be through a server, virtual server or a containerized server.

**Peer** - the entity running a node or set of nodes. In most cases a node and peer are synonyms, with the node being a particular part of a peer's architecture while interacting with other peers.

**Private data manager** - The software that manages private transactions and data. This is the software that runs on the node, a module/segment of the software that runs on the node. In the Ethereum ecosystem this is commonly referred to as a private transaction manager.

**Contract** - A program that runs within the node software that manages the business logic of the blockchain application. This is usually run in a deterministic environment and executes in consensus with other nodes in a network.

**Blockchain application** - a system powered by software used to solve the problems of its stakeholders (companies, governments, citizens or organizations). This system uses the blockchain solution to improve data integrity while maintaining privacy. "Dapps" is a commonly used term in industry to describe blockchain applications.

**Signature policy** - which actors are required to sign a transaction for it to be deemed valid. Signature policies for transactions can also be inherited from parent policies and groups of policies.

## 2.2 Hyperledger Fabric v2.3.1

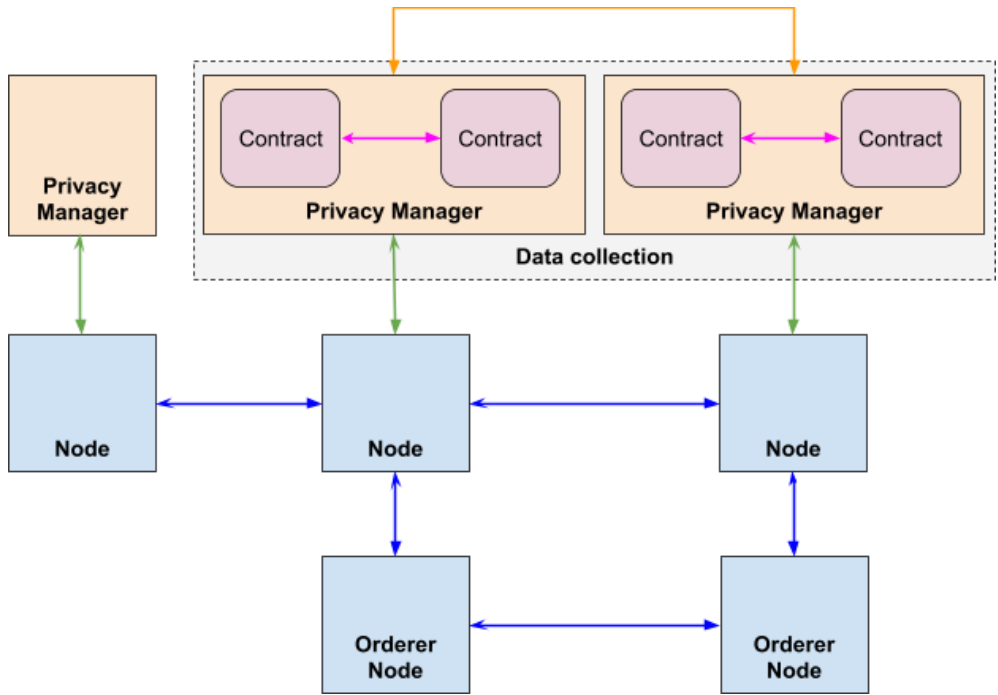

*Figure 1: Hyperledger Fabric high level architecture identifying data privacy boundaries*

Hyperledger fabric is an open source software created and maintained by the Linux Foundation and IBM that can deploy and manage private distributed ledgers to facilitate transactions.

Each Fabric network has multiple encapsulated levels to contain data privacy relationship groups:

**Private ledgers "channels"** - Each Fabric network can have multiple channels, each with their own separate private blockchain ledger. The ledger and its transactions are visible to only other members of the channel or, if desired, can be made visible to all members of the fabric network. Each channel has a governance contract which manages the permissions to read/write data to the ledger through transactions, as well as permissions to add and remove peers from the channel. Contracts deployed at a channel level are only visible to all members of the channel.

Each channel consists of authorized organizations which have two types of peers running different types of node:
1. Anchors peers: The first point of contact for new nodes joining the network to receive network topography information necessary to contact other nodes.
2. Network peers: Peers in the network using one or more channels together. Network peers interact through contract "chaincodes" they install on their nodes which execute the business logic of the blockchain application. This peer will sign the transaction with its own public key (X509 certificate) at the time of endorse transactions.

Every peer holds the current world state and the blockchain ledger for each contract that they manage (if they have it). So, if multiple chaincode exist in the channel then multiple pairs of (world state and ledger) would be present peers that host multiple contracts.

**"Private data collections"** - Each channel can optionally have multiple data privacy collections. Each organization in a channel manages a private database with a segment for each data collection the peer is part of. Each data collection has a governance contract which manages the permissions to read/write data from the private database through transactions of contracts in the data collection, and to add and remove organization from the data collection. Contracts deployed at the data collection level are visible to all organizations in the data collection. Data visibility is granted at an organizational level, not a peer level.

Private data channels consist of an additional two data structures named "private state" and "transient state". Therefore, a peer that is a participant of private data collection will have a world state, ledger, private state, and transient state. Private data objects masked as 'transient', i.e. private, will only be written in the private state and temporarily in the transient state (from endorser phase to commit phase of transaction). Orderer and client SDK will not receive read/write data set transactions of private collection. Only transaction hash will be communicated during the cycle. [1]

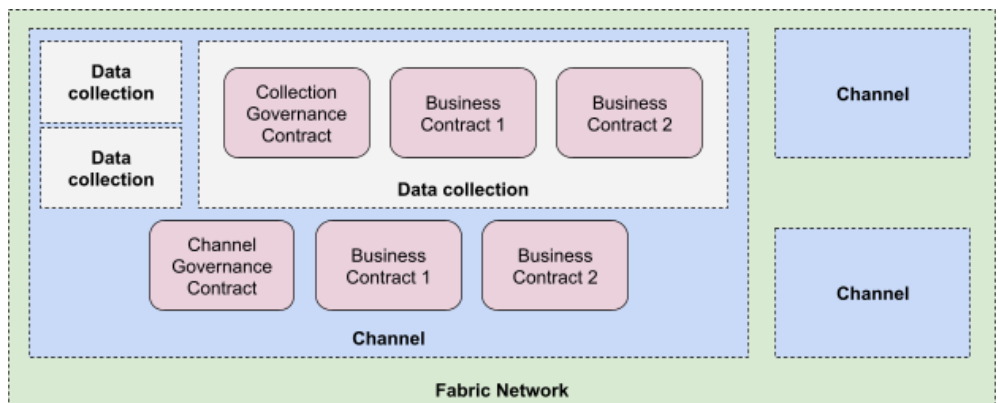

*Figure 2: Hyperledger Fabric network topography*

There are a few types of nodes in a channel:
1. Channel peers - Do business together using a shared set of contracts. Peers manage the channel using the governance contract and manage

any data collections and respective private contracts and state. There are three types of peers as mentioned previously.
   2. Orderers - A special node who receives transactions and puts them into an order in a block. They are responsible for ordering transactions in consensus with other ordering peers, and then distributing back to the peers in the network for committing the transaction in ledger via ledger peer.

Data privacy can be defined at a channel or collection level. Both have a governance contract to manage privacy and permissions as well as business logic contracts to manage state for data for the business operations between the peers of that group. Transaction privacy rules are governed by the executed application contract as well as governance contracts from the collection, channel and network.

### 2.2.1 Contracts

Contracts "chaincodes" are used to manage the business logic of channels and data collections, as well as to define the access control rules and data application business logic. Contracts follow an account based data model in which persistent data is scoped to the contract so that only the contract can read and write data from its own data space. Contracts can call other contracts in the same channel or data collections they have been granted permission to, which enables them the ability to read/write to other contracts data. If granted permissions, contracts on one channel can read and write to a contract on a different channel only via authorized organization part of both channels. [18]

Contracts are only installed on authorized peers in the channel. This is done through an encrypted TLS connection directly with the authorized peers. In this way business logic (code) is not visible to peers not involved in the business transactions of the channel.

### 2.2.2 Channel transactions

Transactions on channels are distributed directly to peers that are required to sign the transaction. Peers are required to sign the transaction as defined in the "endorsement policy" of the contract. The contract may also inherit endorsement policies from the data collection, channel and/or network. The peer executes the transaction and ensures it does not violate any of the contract's logic and adds a signature to it. Once the required number of signatures are collected, the original transaction proposer sends this transaction to an order peer. Call data between peers is sent using an encrypted TLS connection.

The orderer peer puts the transaction in a set order in a block, comes to consensus on that block with other orderer peers, and then distributes the block back to all peers in the channel. All peers in the network receive and validate the blocks.

Peers only have contract code for business operations they are involved in and they can execute transactions if they have the appropriate contracts.

Transactions added to the ledger contain a hash of transaction inputs, and a hash of the key and value for each item in the world state that the transaction reads, writes or deletes (called the read/write set). Only the peers who endorse the transaction, and other authorized peers have raw transaction inputs and the transaction results shared with them. As the value of read and writes may only be simple data, e.g. 1 or "config", the hash of these values can easily be compared to precompiled hash dictionaries, so it is suggested to add random salt to each key stored in the world state to improve privacy.

With only a hash of transaction inputs, function name and results on chain only the following information can be seen by other peers:
1. Identity of all peers that endorsed the private transaction.
2. Time of transaction
3. Contract name and ID

Within each channel, transaction and data can only be seen by peers of the channel. New members added to a channel can see the entire channel's history and ledger data including previous transactions. Peers removed from the channel are still able to see the ledger history up to when they were removed but cannot see new transactions. [19]

### 2.2.3 Private data collection transactions

Transactions in private data collections have a similar flow to transactions on the channel. Peers of data collections create contracts which specify the rules for how the peers can update the data collection's transactions, and business logic for the blockchain application. The contract is installed directly using an encrypted TLS connection.

Transactions are created and signed by peers within the data collection. However, unlike a normal channel transaction, only a hash of the transaction (not the read/write set) is sent to the channel via an Order node. This hash is then ordered by the orderer and put into a block. The peers of the private data collection can recognize the hash and validate the transaction from transient store data structure, and it is updated to a private state data structure of the peer.

Transaction flow for private data collection:
1. Client SDK creates and sends a transaction proposal to all endorsing peers for execution.
2. Endorsing peers will save the executed data in "transient state" data structure for further validation.

3. Endorsing peers will disseminate the read/write set to another authorized private data collection peer.
4. Client SDK will receive the transaction with just a transaction hash (not read/write set).
5. The same hash will be forwarded to the orderer.
6. The orderer will send the hash to all organizations who are part of that private data collection to commit the transaction.
7. Once committed successfully, peers will validate the consistency of the transaction based on the order and finalize the transaction data state.

The main data privacy advantage of data collections is that the read/write set is not visible to peers on the channel, meaning that the hash dictionary vulnerability does not exist to channel co-peers.

Each member of the data collection within a channel can see all transactions within the data collection. Data collection peers run a private database that tracks the transaction data and state. Only the hash of transactions from data collections is added to the channel ledger. New members added to a data collection can see the entire data collection's history and ledger data. Members can be removed from a data channel by editing the governance contract's endorsement policy. Afterwards they will no longer be able to execute any read/write actions using the channel's chaincode. However, they will be able to see the data up to the last block they already had synced on their peer. In order to remove a peer from a data collection a new data collection needs to be created with the new set of desired peers. [20] [21]

### 2.2.4 Data purging

For use cases where private data only needs to be on the ledger until it can be replicated into an off-chain database, it is possible to "purge" the data after a certain set number of blocks, leaving behind only a hash of the data that serves as immutable evidence of the transaction.

There may be private data including personal or confidential information that the transacting parties do not want disclosed to other organizations on the channel. Thus, it has a limited lifespan, and can be purged after existing unchanged on the blockchain for a designated number of blocks using the "blockToLive" property in the collection definition. [22]

### 2.2.5 Protecting private data content from brute force privacy mining

If the private data is simple and predictable (e.g. transaction dollar amount), channel members who are not authorized to the private data collection could try to guess the content of the private data via brute force hashing of the domain space, in hopes of finding a match with the private data hash on the chain. It is suggested that private data that is predictable include a random "salt" that is concatenated with the private data key and included in the private data value, so that a matching hash cannot realistically be found via brute force. The random "salt" can be generated at the client side (e.g. by sampling a secure pseudo-random source) and then passed along with the private data in the transient field at the time of chaincode invocation. [23]

### 2.2.6   Who sees what?

1. **Channel peers** can see the identity of all peers and see the hash of transaction inputs and read/write set hashes for all transactions they have not been shared.
    a. Transaction participants can see the raw inputs and read/write set and receive the contract code to execute and validate this data.
    b. Non-participants can see the identity and transaction's hashed metadata: hash of inputs and read/write set. For data collection transactions the transaction metadata is not visible.
2. **Channel anchor peer** can see the same as non-participating channel peers.
3. **Data collection peers** can see data for all transactions in the data collection which they are participating in.
4. **Orderer peers** do not have any contract code, so only sees transaction input hashes and outputs but does not execute any transactions and cannot see transaction results.

[23] [24] [25]

## 2.3 Corda v4.7

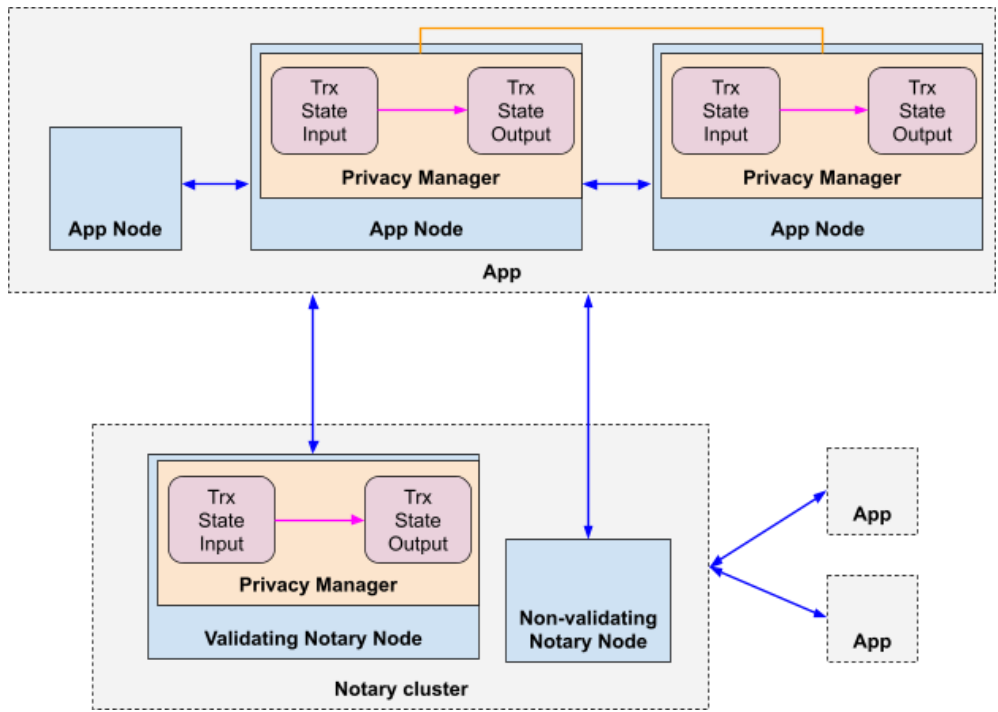

Figure 3: Architecture diagram of a Corda network "zone" identifying data sharing boundaries

Corda is an open source distributed ledger technology created and maintained by R3 that can be used to deploy networks in which multiple peers come to a distributed agreement on the order of transactions which are only visible to participants of the transaction.

There are currently two global Corda networks, each with different operator membership models and transaction fees. Blockchain applications can also deploy and manage their own Corda network. The Corda Enterprise Network Manager tool can also be used to set up a new private Corda network. [26]
1. The Corda Network - The official network supported by R3.
2. Corda Lite - A community led network.

Within the network there are several types of peers who only synchronise parts of the transaction ledger according to the data privacy policy and what applications they are involved in. Within the network, it is possible that no nodes will see the entire transaction ledger. There is no global consensus model because there is no global transaction ledger within the network.

Blockchain applications create an application "CordApp" on the ledger. Each app is run by a set of permission peers and has a set of contracts that define the rules, permissions and data for transactions. Contracts define the data models "state" such as assets, identities, bonds or other data which the blockchain application wants to track. The contracts use an unspent transaction output "UTXO" model in which each new transaction must reference an existing transaction that has not yet been referenced as "spent". The contract code defines the rules for who can create a new transaction that spends the outputs. Unspent outputs of each transaction are the state which is tracked by the permissioned peers.

Only the peers that are involved in the transaction can view the transaction on the ledger and its data. Included peers keep a private database of the unspent transaction's current state in a private database. This means that only the peers that spend the transaction inputs (the ones who create and sign the transaction) and peers that are referenced in the outputs can see the data. Other peers in the app that are not referenced in the transaction cannot see any transaction data.

Each Corda network has a cluster of notaries that ensure the correct ordering and consistency of the transaction states (prevention of an unspent transaction output to be consumed twice). The purpose of the notaries is to ensure that all peers can agree on the order of transactions and ensure that transactions follow the logic defined in their contracts. There are two types of notaries:
1. Non-validating notary - Receives only the hash of the transaction in an application and puts them in a chronological order in the transaction ledger. This peer also checks that none of the referenced transaction inputs have already been spent.
2. Validating notary - Checks both the transaction inputs, as well as that the transaction state changes according to the rules defined in its contract. This means the validating notary must see all of the data in all transactions in each app.

Each application must choose at least one non-validating notary. Validating notaries are not required but suggested for fully trusted transactions. Each state used in an application's contract can be tracked by a different validating notary to reduce exposure of one notary to all types of data. If the transaction uses multiple types of states, then all notaries for each of the states must sign the transaction. Applications can and often use only one notary but are also able to choose several notaries that are then required to come to a distributed consensus using a chosen private consensus algorithm. This can be done to provide greater data integrity guarantees, and to ensure availability of notary services at any time.

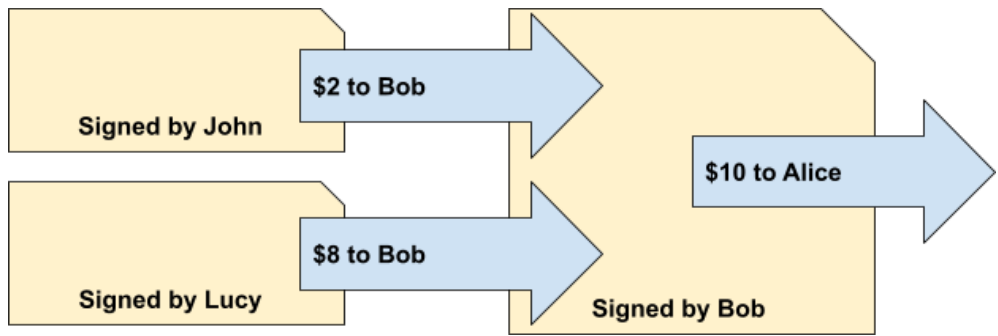
*Figure 4: Corda UTXO transaction example*

As an example, Bob wants to send $10 to Alice. He creates a transaction that references two unspent transaction outputs: $2 he received from John and $8 he received from Lucy. Bob then creates an output in which $10 is referenced to Alice. This transaction is signed by Bob and sent to a validating notary and to Alice. The notary will then check the rules of the transaction (defined by its contract) which say that the sum of money inputs $2 + $8 equals the sum of money outputs $10. After validating the transaction, the notary adds it to their ledger and confirms to Alice and Bob. Alice and Bob validate the transaction's integrity on their ledger and update their private database of their transactions. Important for data privacy, Alice can see in the transaction outputs that John and Lucy both sent money to Bob even though Alice was not part of those transactions.

When a transaction requires multiple peers to authorize it according to the application's signature policy, one of the parties creates the initial transactions and sends it directly to other parties to sign through an encrypted TLS connection. Once all signatures are collected, the transaction is sent to the notaries.

### 2.3.1 Who sees what?

1. **Application peers** can see data of all transactions that they are an input or output of. This includes visibility of all the outputs that a transaction references even though they may not be relevant for the transaction outputs. If a new peer is added to an application, they are not able to see transaction data states from previous transactions unless they later participate in a transaction that references previous transactions.
2. **Non-validating notaries** can see hashes of transactions, and which transaction outputs they reference.
3. **Validating notaries** can see all the data and participants of applications that use them.

[27] [28]

## 2.4 Hyperledger Besu v21.1.0

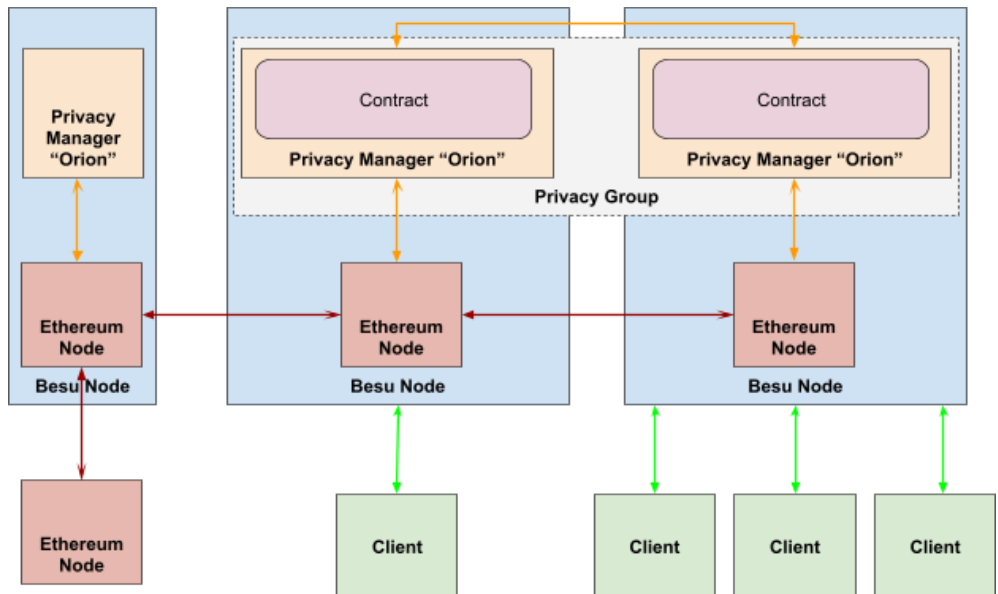

*Figure 5: Architecture diagram of a Orion network which runs on top of the Ethereum mainnet identifying data sharing boundaries*

Hyperledger Besu is an open source Ethereum client, created by PegaSys and maintained by the Linux Foundation, that runs a private data manager called "Orion" and an Ethereum node [29]. Orion manages transactions that are only shared between relevant peers, and a hash of the transaction is put on an Ethereum blockchain network of choice so that peers can verify the transaction ordering and integrity from a shared and trusted source. The private data manager complies with the Ethereum Enterprise Alliance standards.

Besu can be configured to use Clique and IBFT 2.0 Ethereum clients with a proof-of-authority permission network (public or private) or use existing Ethereum networks [30]. For Besu networks intended to be used for private transactions the use of the Ethereum public network is currently not suggested due to its probabilistic block finality property [31].

Groups of nodes running the Orion software can form privacy groups which are then able to see all transactions within the group. Nodes that are not within the privacy group are not able to see the private transactions or the list of participants [32]. However, since every private transaction has a stored pointer in a public chain, the frequency and time of transactions are leaked. Orion also encrypts the stored state of the private transactions. These privacy groups can now be managed on-chain in a smart contract on the shared Ethereum network, allowing additional features to edit the group after it is created and synchronize this between members. [33] On-chain privacy groups are driven by the EEA specifications.

When a transaction is created by Orion, it specifies the privacy group to send the transaction to. The Orion peer then sends the transaction data directly to other peers in the group through an encrypted TLS connection [34]. If peers want to send private transactions between them and are not in a privacy group together then a new one must be formed. Besu's latest version enables the creation of on-chain privacy groups using smart contracts to store, add and remove members from the group, but it is still not recommended for production deployments [35].

Once all members of the privacy group acknowledge that they have received the raw private transaction data, the transaction creator will add only the hash of the transaction onto the shared Ethereum network. Privacy group members can then use the transaction data hash to validate the transaction they received was put on the chain correctly, while other nodes in the Ethereum network are not able to see the transaction data. The transaction on the shared Ethereum network propagates to other peers using an encrypted gossip protocol.

The Besu software keeps track of the Ethereum network shared state as well as the state of all privacy groups that it is involved in. The Orion software manages the private transaction distribution to peers and then sends only the hash back to the Besu node which adds it to the shared Ethereum network. Orion peers use a unique Ethereum key and address for use in private transactions (one for each Orion node). This is different from the address used by Besu on the Ethereum network. The hash of the private transaction that is added to the Ethereum network is signed using the Besu node's Ethereum address, not the Orion address. The mapping of Besu node addresses to Orion node public keys is kept off-chain to conceal their public identity.

An Orion peer is able to create private smart contracts within each privacy group. Private contracts use the same account and state model used by Ethereum [36]. A contract in a privacy group can read data from contracts in the same privacy group and from the shared Ethereum network contracts. A contract in a privacy group cannot access data from contracts in a different privacy group and it cannot modify information in the shared Ethereum network. Contracts on the shared Ethereum network cannot read or write to contracts in any privacy group [33]. A peer in a privacy group can read all contracts deployed in the privacy group. The Orion software automatically creates new privacy groups when contracts are deployed if involved peers do not already exist in a privacy group.

The privacy group identities are detached from the public chain identities. Each Orion node has a key pair for signing private transactions that is not related to the Besu public key. Besu may use a different key to register a marker to each private transaction in the public chain, therefore achieving a certain level of anonymity [37]. However, this becomes less practical when funds must be transferred to new addresses to be able to register transactions. Orion keys, on the other hand, are intended for long term usage, generating trust among peers.

An Orion peer can have several independent clients attached to it called "tenants". These clients use a JWT token to authenticate to Besu and each of them use a separate Orion key [38] [39]. The Besu/Orion node operator has access to the tenant's data and Orion public keys, and it can control what information each tenant has access to depending on which privacy group it belongs to [7]. This feature exists to allow role-based access to the same node. Each client can see the mapping of Besu to Orion addresses [39].

When a transaction is sent between peers in a data privacy group, which is then broadcast to the shared Ethereum network, each member of the data privacy group can figure out the mapping between the Ethereum transactions address and the corresponding Orion address. If the Besu node uses the same Orion address in private transactions with different (other) peers, then the other Besu peers that know the mapping can view basic activity data about private transactions that they are not involved in:
1. Besu identity
2. Timestamp of transactions

### 2.4.1 Who sees what?

1. **Clients (tenants)** use authenticated API requests to Besu, which allows them to see only information of the private groups they are involved in.
2. **Privacy group peers** can see all contracts and all transactions between all peers in the group. Can see data for all clients attached to the peer.
3. **Ethereum mainnet peers** can see transaction hashes and signatures from Besu node's Ethereum network addresses, including timestamp. They cannot see their Orion address used in private transactions unless they have previously been involved in a transaction with their Orion peer.

## 2.5  Quorum v21.1.0

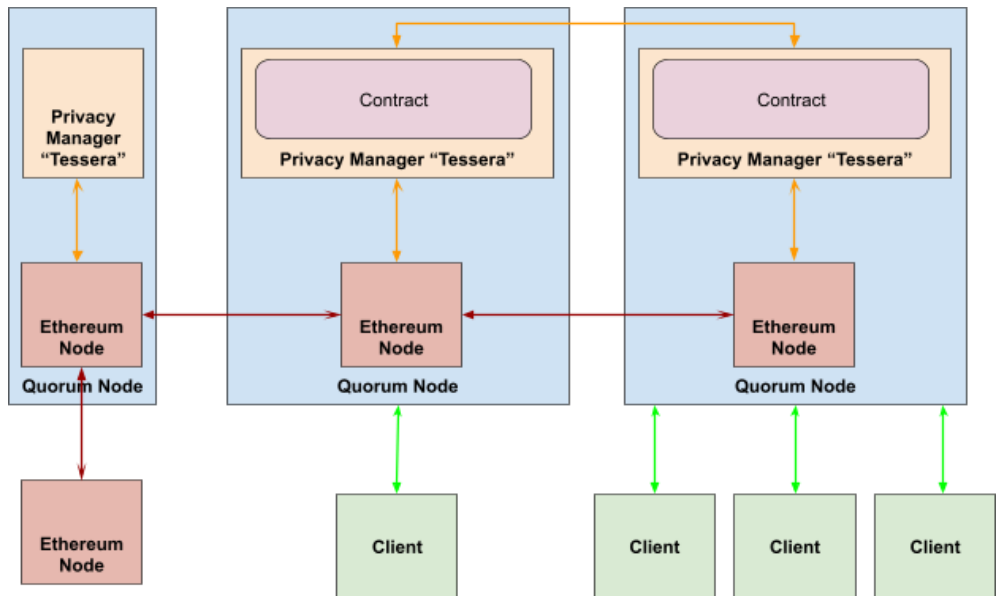

*Figure 6: Architecture diagram of a Quorum network identifying data sharing boundaries*

Quorum is an Ethereum-based distributed ledger protocol with transaction and contract privacy and new consensus mechanisms. Quorum is released as open source software created and maintained by Consensys who recently acquired it from J.P Morgan in 2020. Quorum node consists of an Ethereum-based blockchain client and a privacy transaction manager. Tessera is one of several transaction managers that work with Quorum. Tessera manages private transactions that are only shared between relevant peers. A hash of the transaction is put on an Ethereum blockchain of choice so that peers can verify the transaction's ordering and integrity from a shared and trusted source. The project also complies with the Ethereum Enterprise Alliance standards. [40]

Quorum has a similar architecture and transaction flow to Hyperledger Besu. The key differences lie in how they mark and order their respective private transactions as well as choice of consensus algorithms.

Quorum is usually used to run private networks. Quorum can operate in a quasi-public mode with IBFT consensus -- in this set up, validator nodes would need to be set up in advance, but reader nodes can be added in at any time. Quorum cannot be run with the Ethereum public network due to its probabilistic block finality property.

The private recipients to send the transaction to are specified when a private transaction is created. The private transaction manager then sends the transaction data directly to the recipients through an encrypted TLS connection.

Once all recipients have acknowledged that they have received the raw private transaction data, the transaction creator will add only the hash of the transaction onto the shared Ethereum network. Peers party to the transaction can then use the transaction data hash to validate the transaction they received was put on the chain correctly and execute it to update the nodes state, while non-participating nodes in the Ethereum network are not able to see the transaction data. The transaction on the shared Ethereum network propagates to other peers using a gossip protocol. [41]

This is slightly different to Besu, which can alternatively use a dedicated privacy group as the recipient of a private transaction, instead of only a using list of peers like Quorum. The data privacy properties of both mechanisms are the same.

The Quorum node keeps track of the Ethereum network public and private state. The private transaction manager (Tessera) manages the private transaction distribution to peers and sends only the hash back to be added to the Quorum node.

Tessera peers use a unique key and address for use in private transactions (Tessera can host multiple private keys, but at least one is required to be able to send private transactions to a Quorum node). This is different from an address used by the Quorum node. The hash of the private transaction that is added to the Ethereum network is signed using the Quorum node's address, not the Tessera address. The mapping of Quorum node addresses to Tessera node public keys is off-chain to conceal their public identity.

A Quorum peer can have several independent clients attached to it called "tenants". These clients use private keys to authenticate to Quorum. This feature exists to allow role-based access to the same node. Each client can see the mapping of public keys to the Quorum address. This is called "Multi-tenancy". [42]

When a private transaction is sent between Tessera peers, that is then broadcast to the shared Ethereum network, the recipients of the private transaction can see the mapping between the Quorum address and the corresponding Tessera address. If the Quorum node uses the same Tessera address in private transactions with different (other) peers, then the other Quorum peers that know the mapping can view basic activity data about private transactions that they are not involved in [43]:
1. Quorum identity
2. Timestamp of transactions

### 2.5.1 Who sees what?

1. **Clients (tenants)** use authenticated API requests to Quorum, which allows them to see only information of their role (scope) allows.
2. **Quorum transaction recipients** can see transactions sent to them.
3. **Quorum peers** can see transaction hashes and signatures from Quorum node's Ethereum network addresses, including timestamp. They cannot see their Tessera address used in private transactions unless they have previously been involved in a transaction with that Tessera peer.

## 2.6 Nightfall v1.0.2 (Zero knowledge proofs)

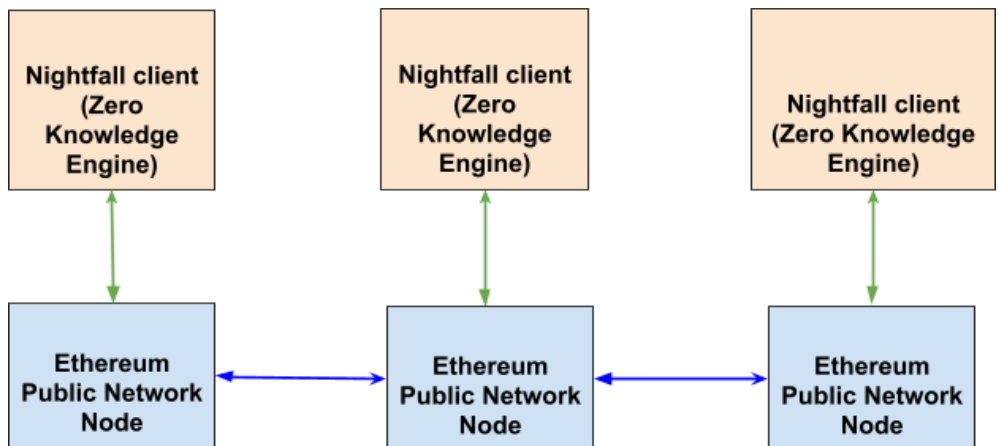

*Figure 7: Architecture diagram for nightfall identifying data sharing boundaries*

Nightfall is an open source software created by Ernst and Young that can deploy and manage transactions of ERC20 and ERC721 token smart contracts that run private transactions through a zero knowledge engine on the Ethereum public network. The software is in an incubation stage and Ernst and Young are testing its usage with their clients.

Unlike a regular Ethereum smart contract, the Nightfall contract uses Ethereum's ZK-SNARKS cryptographic primitives to prove the correctness of the contract's state and transaction validity. This uses a special type of cryptography called zero knowledge proofs, in which a computer constructs a proof (piece of data) that certain logic conditions were met during a program execution. The proof itself does not contain the data used during the program execution. This means that the proof shows validity and data integrity, without revealing the data itself. The data of the proof can be read only by parties holding specific private data.

Nightfall token contracts used a three-step process to create private transactions.
1. The Ethereum user sends any ERC20 token or Ether to the Nightfall smart contract on the Ethereum public network. The user also sends a

zero knowledge proof for a new minted token of the same amount. The smart contract will lock the tokens in escrow and at the same time **mints** a new private token commitment. The private token commitment is an unspent transaction output that references a public key which can spend the UTXO built under zero knowledge such that the amount, token type, senders and recipients are not revealed. The keys used in the commitments are different to the keys used on the Ethereum public network.
2. The user with the private key corresponding to a private commitment transaction outputs can construct a new commitment that will invalidate the old commitment and create a new private transaction with new transaction output recipients. This mechanism acts as a private token **transfer** from one user to another.
3. A user can **burn** a token commitment which will release the ERC20 or Ether from the Nightfall contract to a supplied address.

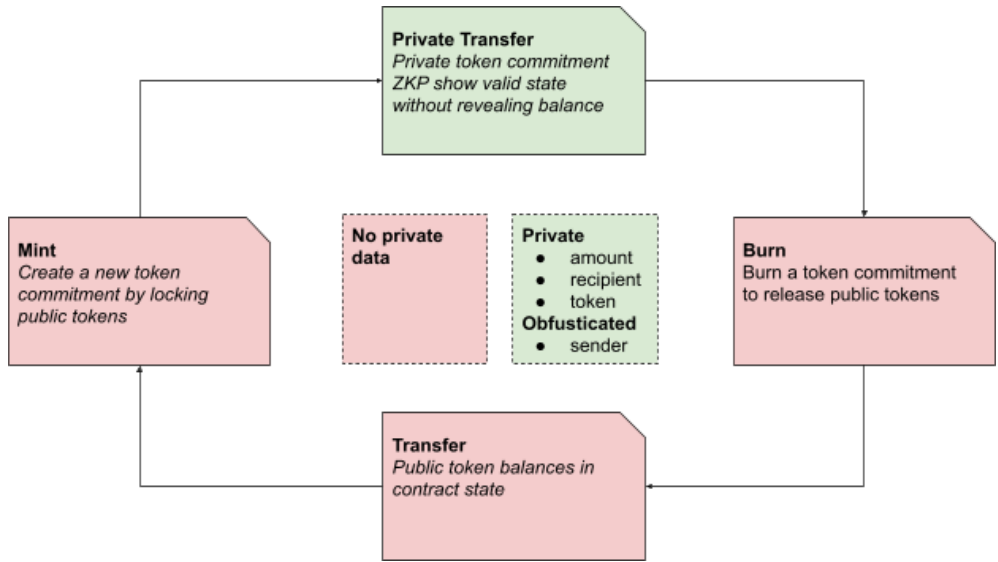

Figure 8: Token transfer cycle and privacy

The key pairs to control private tokens are different to those used on the Ethereum public network. The private transactions use zero knowledge proof commitments to ensure the correctness of these transfers without revealing the amounts, token type or the sender or recipient keys.

The mint and burn transactions do not have any privacy and other Ethereum users are able to see which accounts created or destroyed how many private tokens.

The private transfer transaction still needs to be executed by an Ethereum public network account through a public transaction. It is reasonable to assume that the same Ethereum account controls the key pair used for the zero knowledge commitments in many cases. In this way the Ethereum account sender is also the sender of the private token transfer, revealing the identity of the private sender. If the user wants greater privacy, they can use multiple Ethereum accounts to control the private token transfers. However, each of the Ethereum accounts needs Ether to pay for gas and there are various mechanisms to trace Ether transfers back to the original account. In this way the Nightfall contract makes it much harder to link the identity of the private token transfers but not completely. To truly guarantee privacy an additional data privacy strategy is needed such as the use of a third-party custodian mixer or Ethereum accounts funded using off-chain transactions.

Each private transfer commitment contains a random number generated by the sender. This random number must be transported privately to the recipient for them to unlock the transaction output and spend the token commitment. The whisper network, an underlying private and anonymous data transport layer for the Ethereum application ecosystem, is used for the private data transport between the sender and recipient. If the recipient is not online within the time to live of the sent data (approximately one day) and does not receive this random number through another channel, then the token commitments can never be spent. [44]

Zero knowledge applications comprise two pieces of software. One software creates a zero knowledge proof, which for Nightfall is done by the client software (a nodejs application). The other software verifies the correctness of the proofs, which for nightfall is done within the public Ethereum smart contract. A proofing and verification key are needed for both softwares to work together. These mathematically related keys must be generated specifically for each zero knowledge application. During the generation of these keys some extra data is created. This extra data, called toxic waste, can be used to generate fake zero knowledge proofs in the application. Therefore, users of the zero knowledge application must trust that this extra data is deleted by the people that generate the proofing and verification keys. It is for this reason that the generation of these keys is usually done in a multi-party computation ceremony [45]. In this ceremony multiple people (the amount is limited by the practicality of coordinating such a key ceremony, with 90 participants in the largest recorded) are involved in the generation of the proofing and verification key and as long as one of the people deletes their data, none of the parties are able to correctly generate the toxic waste data that can be used to create false proofs. There are variations of zero knowledge proofs that do not require a trusted setup as described above, with different efficiency and other characteristics. [46]

Zero knowledge proofs are a new and heavily researched area of cryptography. The construction of secure knowledge proof engines is complex and difficult. The nightfall team used the ZoKrates tool [47] to generate the zero knowledge software required, greatly simplifying their development cycle. The fast pace of innovation in this cryptography makes it more susceptible to risks that the current implementation can become redundant or deprecated.

### 2.6.1 Who sees what?

1. **Nightfall contract deployers** are never able to see private token commitments of other users.
2. **Nightfall contract users** can see transaction data relevant to their zero knowledge transaction commitments that they create, and the proofs that send them new tokens.
3. **The Ethereum public network users** can see Ethereum accounts that execute the transfer transactions but are not able to see the key identities used for the sender and recipient of the transfer. They are also not able to view the amount, or the token type transferred. It is reasonable to assume that the Ethereum account that makes the transfer also is the owner of a key that sends private tokens creating an obfuscated token sender.

# 3 Existing EOSIO Privacy Solutions

This section will outline the ways in which the core software manages permissions to create data control and privacy. It will also cover additional data privacy solutions and services that already exist which are powered by EOSIO. These solutions are built by members and companies in the EOSIO community that range from proof of concepts to mature and managed products and services.

## 3.1 EOSIO v2.1

EOSIO is an open source software created and maintained by Block One that can deploy and manage public or private blockchains that support smart contracts. Each blockchain network deployed shares a blockchain ledger that has consistent consensus amongst up to 125 block producers. The governance of the blockchain (default is proof of authority), is defined through smart contracts.

There are several data boundaries in the software at which data privacy can be applied.

Figure 9 shows the different boundaries between data that exist for an EOSIO blockchain. At each of these boundaries, data privacy controls can be enforced with different levels of data integrity guarantees.

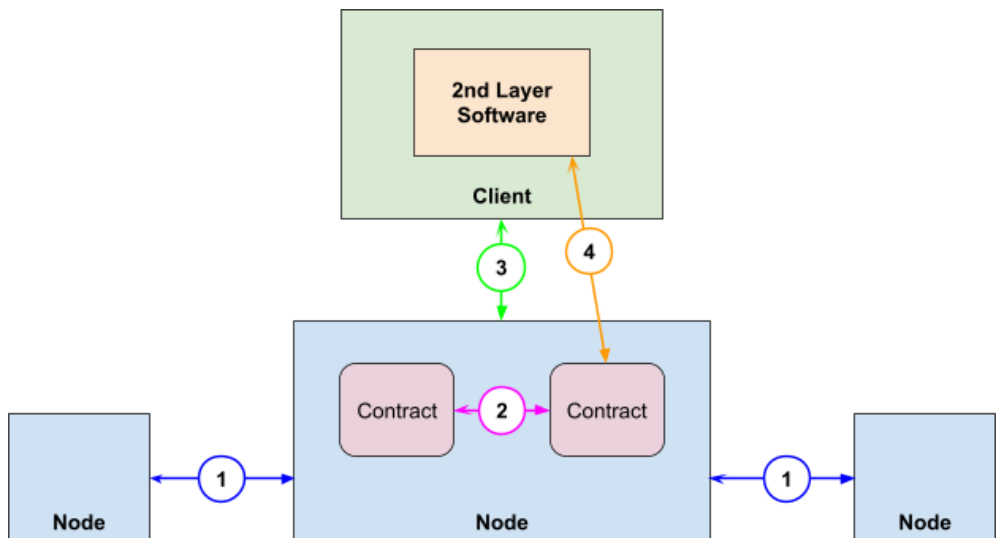
*Figure 9: Data relationship boundaries within an EOSIO blockchain*

### 3.1.1 Data Boundary 1: EOSIO protocol peers

Each node in the network synchronizes its blockchain state and history with other nodes using the EOSIO protocol. Data is sent between nodes using TCP (SSL encrypted or unencrypted). Each node in the network can configure a list of other nodes which are authorized to read and synchronize data. This configuration can be done in the node software using the allowed-connection configuration in the net plugin. [48] Network configuration such as firewalls can be used instead of or in addition to create node synchronization access and encryption.

At the protocol level anyone who synchronizes a node can see all the blockchain data. Permission to synchronize the blockchain is controlled through the net plugin or via firewalls which will lead to private blockchains discussed in Section 3.2.

### 3.1.2 Data boundary 2: Smart contract data

Each smart contract on the blockchain defines the data that it stores in the EOSIO database (a persistent storage space for smart contracts stored on all nodes). All smart contracts can read all other smart contract's data on the blockchain. Read permissions do not exist at the protocol level for smart contracts.

At the smart contract level there is no private data, all contracts can read data of all other contracts.

### 3.1.3 Data boundary 3: Client-server

Each EOSIO node can optionally run an http API which allows connected clients to read data from and send transactions to the blockchain. The HTTP plugin manages connections to clients. [49]

Data privacy can be built for clients accessing the blockchain using an extension of the HTTP plugin, or a middleware. This is not provided by the EOSIO software and the read privacy permission logic provided depends on implementation. Unless the blockchain is run by one entity, all parties must trust that all other parties in the network run the same authentication software on their client API to ensure data privacy.

Several community APIs exist with enhanced features:
1. Dfuse [50]
2. Hyperion [51]
3. EOSIO Light API [52]

### 3.1.4 Data boundary 4: 2nd layer software

Additional data privacy techniques can be built using smart contracts with a second layer of software in the client. These utilize cryptographic data structures such as hashes and encryption that put a concealed version of the data on the blockchain, while the unconcealed version stays private on the client managed by the privacy enabling software. One or more smart contracts manage the data on-chain and a second layer software manages the private data in the client. The concealed data on-chain is authorized using an account's permissions, while the second layer software manages all the off-chain private data. [53] This is the same mechanism use by Nightfall on the Ethereum public blockchain.

The data privacy and integrity of second layer software solutions depends on the implementation. Section 3.3 will discuss second layer solutions in more detail.

Table 2 summarizes the different data boundary permissions, what is responsible for managing the permission, and whether it falls into the category of being managed through smart contracts on the blockchain. Permissions that are controlled through smart contracts inherit the data integrity properties of a blockchain while other data permissions are subjective and can only be enforced on a per-node basis without cryptographic guarantees. Data on the blockchain (raw or in a concealed form) provides the data with the level of cryptographic integrity and verification that is not provided in current database systems while the data remains private.

### 3.1.5 Summary

| Data boundary | Permission management by | High data integrity |
| --- | --- | --- |
| 1: EOSIO peers | Net plugin | |
| 2: smart contract data | **Not currently supported by EOSIO | Yes |
| 3: client-server | Http plugin | |
| 4: 2nd software | Smart contract + 2nd layer software | Yes, as implemented in smart contract |

Table 2: Summary of EOSIO data privacy boundaries

The EOSIO software is in stable production use on at least a dozen public chains and over a hundred private blockchains. There has been large adoption of the software with one of the blockchain industry's largest communities. Each public blockchain has their own subset of community supporters, tools, philosophies and users. Official documentation is maintained by Block One and the community has expanded on this with a vast number of articles, tools and guides.

The EOSIO software and default governance contracts provide a highly scalable proof of authority blockchain that can be tailored to many stakeholders needs. The software does not yet support data privacy at a protocol level, a variety of mechanisms exist to manage private data using the integrity of EOSIO technology [54] which are discussed below.

## 3.2 Private EOSIO chains

A completely private EOSIO chain is not able to be read or accessed except by authorized peers. A private chain can be created by setting a few configuration variables in the http and net plugins for each node in the network as explained in Section 3.1.1. Different configurations are possible allowing for privacy between nodes running the blockchain as well as between the blockchain nodes and connected client software.

Instead of using the features of the EOSIO software, firewalls, VPNs and other network access controls can be used to run an EOSIO blockchain in a private environment as well.

The authors of this report are aware of over 100 different entities with private EOSIO chains at various stages of development. Unlike public EOSIO chains which are inherently public, these chains cannot be identified except for public releases of their existence. The following are some examples with public information:
1. SM Jaleel & Company [55]
2. Europechain [56]
3. Grant Thornton [57]

Other products and services exist and use private EOSIO blockchain technology outlined below.

### 3.2.1 Public or private blockchain as a service

Strongblock and Dfuse offer a blockchain as a service platform with the ability to deploy bespoke blockchains including private blockchains. Blockchain nodes deployed through the Dfuse platform are additionally powered with the Dfuse API software, an open source API designed for faster and easier data queries and other features. [50] [58]

### 3.2.2 Public blockchain supporting GDPR data privacy

Europechain is a permissioned public blockchain powered by EOSIO. The blockchain has been configured to support GDPR compliant data privacy for enterprise needs by adding a layer of legal and technical measurements. Accounts can request permissions through a portal and sign compliance to the terms and conditions. Access to the history is regulated by nodes that have implement filters for the accounts that have requested to be forgotten. The blockchain nodes are configured to only synchronize with each other so that no additional parties can access the "forgotten" data.

Each account can signal to the blockchain if it would like its account to be forgotten. This is done by sending a transaction to a system contract, which will cause the history nodes to withhold any data about that account's state or history if anyone requests it. [59]

Applications need to provide a privacy impact assessment as is required in the GDPR regulations.

The chain's governance consists of a consortium of the founding 5 block producers plus the additionally selected block producers, all are based in the European zone. Additional Block Producers are IT firms with 7*24 hours operations and a strong brand.

### 3.2.3 Private databases pegged to a public chain

A common criticism of private blockchains is the inability for external parties to verify or audit the blockchain in full or in part, which reduces the value added by using a blockchain data structure and synchronization protocol. There is a simple technique to bring back a lot of this value which is to place a summary of the private blockchain in the form of a hash on a public decentralized blockchain. With this technique, external parties get much stronger data integrity guarantees about data from private blockchains. This additionally adds extra security to the private blockchain by ensuring that misconduct in block consensus can be more easily detected.

This technique can easily be done with EOSIO private blockchains by putting the block headers from the private chain onto a public EOSIO chain.

Blockbase has built a blockchain powered database (not using EOSIO technology) in which the block headers are put on the EOS or Telos public decentralized EOSIO blockchains. [60]

## 3.3 2nd layer software

EOSIO smart contracts have built-in efficient support for several widely used cryptographic primitives such as hashes. EOSIO smart contracts can also use code from well-maintained c++ cryptography libraries. The flexible programmability of EOSIO smart contracts allows for a wide variety of second layer software techniques to exist.

### 3.3.1 Private data communication

Sending a private message between two blockchain accounts can be done in a variety of ways with different characteristics.

**Encrypt data**

This strategy involves encrypting data using the recipient's public key obtained from the blockchain and putting the encrypted data directly into a smart contract. The recipient can then read the encrypted data and decrypt it. The strategy is simple and depends on no external infrastructure. The time of the message and the recipient are not private. There are also additional concerns that the encrypted data, which is available to anyone who can access the blockchain, can be decrypted later when quantum computing matures, or techniques mature.

Both implementations below use symmetric encryption for the data with a shared key derived from the recipient's public key and the sender's private key (using the Diffie–Hellman key exchange [61]). Using symmetric encryption reduces the amount of data in the encrypted message. As storage capacity is limited on the blockchain, sending large amounts of data using this method is infeasible even when using symmetric encryption.

| Name | Status | Notes |
| --- | --- | --- |
| Pobox [62] | Proof of concept | Uses a custom smart contract to send messages |
| EOS communication [63] | Proof of concept | Sends the encrypted message in the memo of a token transfer |

Table 3: Software to send private data communication using encrypted data

### Encrypt data hash

This strategy involves encrypting data and then putting a hash of the encrypted data on the blockchain. The encrypted data is stored using external infrastructure that both parties can access. The recipient looks for messages addressed to it on the blockchain, retrieves the data from the external infrastructure and verifies that the data hash matches the on-chain hash and then decrypts it.

This can be done using public infrastructure such as IPFS or a server. IPFS uses existing shared infrastructure making it easily accessible. This may also have concerns that the encrypted data can be decrypted later when quantum computing matures. IPFS has a very high availability but can have latency issues when first requesting data that is far away. Using centralized controlled servers such as AWS S3 buckets or a database gives full control of access and availability to the server.

### Chappe

The sender and recipient can be private by publishing a private key so all parties may publish to the network using the same EOSIO account, obscuring identity. The recipients may not know how many or who have access to the channels' private keys. This enables metadata obliviousness. The EOSIO network and time of the message, however, cannot be private.

Chappe also supports read receipts. In this case, recipients of messages decrypt the content, sign it with their device key, and then encrypt and re-publish that signature. This allows any participant on that channel to prove that the party sending the receipt successfully acknowledged the contents of the message. [64]

| Name | Status | Notes |
|---|---|---|
| Chappe [65] | Proof of concept | Uses a IPFS to store encrypted message data |

Table 4: Software to send private data communication using an encrypted data hash

**Encrypted connection handshake**

This strategy involves encrypting the handshake data needed to create a secure internet connection between two parties, brokered through a smart contract. This may involve multiple transactions from both parties to broker the handshake. This strategy can use TCP, webRPC or any other web communication protocol to send data. Once the communication channel has been brokered through the smart contract, the two parties can send data directly to each other without any interaction on the blockchain.

To send data to a recipient requires that they are online to be able to broker the communication handshake. Once the connection exists blockchain infrastructure is no longer needed and as much data as the communication protocol supports can be sent. Using a protocol such as webRTC allows the two parties to stream videos in real time and send other media formats supported by webRTC. This is not possible if data is sent through the blockchain. [66]

None of the message data, even in encrypted or hashed form, is written to the blockchain making data privacy using this method significantly higher and making it scale to much larger data volumes.

| Name | Status | Notes |
|---|---|---|
| Sense Chat [67] | Production app with significant adoption | Sends messages via the EOS blockchain. If the recipient is not online, then a message is sent using an end-to-end messaging service without any interaction on the blockchain. Brokers a webRTC connection using a smart contract. |
| Funnels [68] | Proof of concept | Brokers a webRTC connection using an EOSIO smart contract. |

Table 5: Software to send private data communication using an encrypted connection handshake

**Summary of private data communication**

Table 6 shows a summary of the different data communication strategies using EOSIO blockchain technology. It outlines the unencrypted data that is put on the chain and constraints and advantages for the strategy. In all cases the message data is private.

| Name | Unencrypted data | Constraints | Advantages |
|---|---|---|---|
| Encrypted data | Sender, recipient and time and size of message | Blockchain storage capacity cannot handle significant message data volume. | Simple and no external dependencies |
| Encrypt the data hash | Sender, recipient and time of message | | Can scale to much higher data volume |
| Encrypted connection handshake | Sender, recipient and time that communication session started | Both users must be online to send data | Can scale to very high data volume and support multiple data transfer types including streaming |

Table 6: Summary of private data communication strategies using EOSIO technology

### 3.3.2 Private data storage

There are several strategies for storage of private data associated with an EOSIO blockchain account.

**Client side data encryption**

The traditional way to store private data of an EOSIO account is to encrypt it on the client and store it on external infrastructure. For data efficiency it is often best to encrypt the stored data with a symmetric key and encrypt the symmetric key with the private key of the account. Data can be stored on public infrastructure such as IPFS which has high availability but may encounter other issues. Using a trusted data storage such as AWS S3 buckets has additional privacy benefits for the data and offers greater control over access.

**Self-sovereign data**

The use of a person's device to store, and selectively disclose, their own data is known as self-sovereign data. This relatively new concept puts users in primary control of their personal information and can significantly reduce technical and legal risks associated with storing personal information.

There are several challenges associated with self-sovereign data.
1. If a user's device breaks or is lost, then data may need to be recoverable. Standard techniques can be employed to create encrypted backups on third party infrastructure. The Universal Backup System [69] and Encrypted Data Vaults [70] is a solution for this for the self-sovereign identity ecosystem.

2. Data stored exclusively on a user's device can be difficult to query and can only be accessed when the device is online. Third parties can be trusted to provide access to data when a user is off-line or if their data needs to be continually queried.

3. The user's client software is in control of what data it provides when it is requested, and therefore different versions of their sovereign data can be provided depending on what it is requested for. The use of high integrity data as explained in Section 3.3.3 can be used to mitigate this issue.

**PrivEOS**

PrivEOS is a decentralized backup and recovery system. A user can split up data into several pieces using Shamir's secret sharing and distribute the pieces to several nodes in the PrivEOS network. Each node only possesses a single piece of the key and is not able to recover it in full. The user can define the rules under which the key can be recovered. Under normal conditions the user can request all the pieces of the key and construct it themselves.

The system can be used to host high-value information such as private keys or other data in a way that is both private and recoverable. It is not designed for large amounts of data.

| Name | Status | Notes |
| --- | --- | --- |
| Client side data encryption | Many production ready solutions exist | Technique to encrypt data client-side |
| PrivEOS [71] | Ready but not in operation | Decentralized private data storage and recovery |

*Table 7: Software to store private data*

### 3.3.3 Private data with high integrity

Ensuring that another party has seen the data you see and is providing the same data to you as to everybody else is a high value property for private data. Systems that can do this provide necessary guarantees when working with external parties and can additionally make sovereign data viable. Rules over how the data is updated and structured can be cryptographically enforced giving parties privy to that data even stronger control over their mutual business operations.

**Private sovereign smart contracts**

A lightweight private blockchain can be run by each of the parties that need to share data privately. Transactions are distributed only to the parties that need to see the data. This lightweight blockchain need not have all the overhead as a regular EOSIO blockchain and thus only needs to update its state when new transactions exist (as opposed to creating a new block of transactions even if none exist).

This technique has been implemented in Holochain [72].

A hash of the transaction is stored on a shared EOSIO blockchain which allows all the parties to verify what data each of the other parties saw, without making the raw data available for everybody else in the system. This shared blockchain can be a custom private or public EOSIO chain with governance to suit the stakeholders, or an existing EOSIO chain that matches the requirements of the stakeholders. [73]

The following two technologies show two different ways to implement private sovereign smart contracts.

### EOSIO sovereign contract
This executes an EOSIO smart contract in a Google Chrome browser. Most popular internet browsers now support wasm execution environments, making it possible and highly accessible for many devices.

Using the inbuilt wasm execution environment has several challenges:
1. All the libraries to support EOSIO smart contract execution need to be built in Javascript which is a lot of work to build and additionally to keep up to date with the EOSIO protocol.
2. Wasm import and export functions are executed in Javascript which is asynchronous, making it difficult to replicate the exact same behaviour as EOSIO server nodes. [74] [75]

| Name | Status | Notes |
|---|---|---|
| EOSIO Sovereign Contract [76] | Proof of concept | Run an EOSIO smart contract in the browser |

*Table 8: Software to store private sovereign data*

### 3.3.4 Other 2nd layer solutions

### EOS Blender
This is a web application that can be run in any web browser, connected to a smart contract on the EOS blockchain. Users can deposit tokens into the contract with a hash of the recipient and an unlocking code. These hashes are combined with other users deposits and the recipient can withdraw from the contract without revealing the initial hash, removing the link between depositor and withdrawer. This strategy requires trusting a party to aggregate hashes to create privacy and requires that many people use the service for it to be truly private.

**Peos**

This is a token planned to be launched on the EOS blockchain using similar technologies as Monero [77] to ensure that the sender, recipient and amount are private. The Peos client manages a private account controlled by different keys than the user's EOS account. The user is then able to transfer Peos tokens to their private Peos address and then can make fully private transactions to other Peos addresses. The design depends on several cryptographic primitives which are currently not supported as native wasm execution intrinsic in the EOSIO protocol [78]. Because of this, the execution efficiency is not fast enough for them to execute within the time constraints of the low latency 0.5 second blocks of the EOSIO protocol. The Peos team have written the code to integrate these intrinsics into the EOSIO code base and are currently waiting for the Block One team to approve and integrate the code. [79]

| Name | Status | Notes |
|---|---|---|
| EOS Blender [80] | Operational on EOS blockchain | Requires trust that the centralized EOS blender service will allow funds to be withdrawn and manages private data. Not open source. |
| Peos [81] | Waiting on EOSIO features to complete | |

Table 9: Other 2$^{nd}$ layer solutions

# 4 Analysis Framework

The following framework was used to analyze each of the solutions that were researched:

**Maturity**
What is the status/readiness of the tech? Considers scalability, code stability, complexity and ease of learning/documentation/community.

**Data privacy relationships**
How are data privacy relationships managed? Describe private chains, groups, zones etc...

**Private data transport**
How is data transport (communication) done between peers exchanging private data? E.g. gossip protocol with encrypted data.

**Private data integrity**
How is data integrity maintained? What guarantees do parties get that other parties see and use the same data that they see, and follow the same rules (transaction logic)?

**Data privacy**
Who sees what data? What metadata or private data is leaked and to who?
What happens to entities that are added/deleted to the private relationship after it starts. What data can they see?

See Appendix A: Analysis framework results to see the results of the framework applied to all the reviewed technology.

# 5 Conclusion

A large variety of solutions that provide some level of data privacy while taking advantage of data integrity guarantees provided by blockchain technologies were reviewed. This included mature and well adopted solutions to proof of concepts, and solutions that utilize EOSIO technology and those that build their own or use other types of technologies.

All solutions apply the same general principle: that of keeping raw data off the blockchain and referencing "anchoring" it on the blockchain with a proof to maintain data integrity while keeping data only with relevant parties. The goal is to provide a level of data integrity not provided by a regular database while keeping data private. [82]

These technologies solve various types of data privacy problems including general purpose private business transactions, private high availability data transport, private distributed key management, and restoring and privates token transfers.

The benefits of combining private data with blockchain technology allows enterprise and government to manage data privacy, while providing high data integrity via cryptographic proofs. This allows for systems that are more secure, have higher availability, less incentive for hackers, can provide sovereign data control and reduce data liability. The complexity of the solutions compared to traditional database systems limits adoption. It was found that solutions with data privacy features were more complex than those with less features and flexibility.

None of the reviewed solutions were objectively better than others for all use cases. Each solution provided its own pros and cons. A data privacy solution that works with blockchain technology should be adapted for the applications needs, the use case and other requirements.

## 5.1 Reviewed technologies

### 5.1.1 Private transaction managers

Four solutions were reviewed:
1. Hyperledger Fabric
2. Hyperledger Besu
3. Quorum
4. Corda

Each of these solutions runs a shared transaction ledger which allows peers in the network to come to a distributed consensus on the irreversible order of transactions. All solutions provide some form of a private transaction manager, that executes transactions in a separate process using a separate database with only the hash of the transaction being stored on the shared transaction ledger. The transaction hash gives each of the participants of the private transaction cryptographic proof of its existence and timestamp allowing the peers to come to a distributed consensus on the private transaction and private database.

Hyperledger Besu and Quorum have a very similar architecture, and both comply with the Ethereum Enterprise Alliance standard for private transaction managers. Private data relationships are specified in Quorum at the transaction level, by specifying private recipients. Besu can also use private recipients or alternatively uses data privacy groups which are automatically created for any new transactions with new participant groups. In this way Besu provides an additional layer of flexibility for data privacy groups.

Both solutions have a highly robust data privacy model and mature code base with significant enterprise adoption. In a very subtle way, participants of a private transaction may be able to correlate activity frequency data about other participants' private transactions that they are not involved in.

Hyperledger Fabric ships with out-of-the-box support for deploying multiple private ledgers, each with a separate ability to contain their own extra-private transactions and have their own private data relationship groups (similar to Besu's privacy group) called data collections. While it is possible to deploy and manage multiple ledgers with Besu, Quorum and Corda, the inbuilt management solution allows fabric clients extra flexibility to manage data privacy and complex data privacy relationships.

All transactions on Fabric ledgers are private, with the transaction inputs and results (read/write set) only visible to participants. Contract code is only shared with participating organizations, meaning transaction validation can only be performed by the same. This provides significant additional data privacy for users of the shared ledgers compared to Besu and Quorum shared ledgers.

Fabric uses a read/write set of database keys-value pairs modified during transactions for peers to come to consensus on the result of transactions. The ledger records a hash of these values making them unrecognizable for other peers not involved in the transactions. If the value is simple, hash dictionaries can be used to guess transaction read/write sets, so it is suggested to add salt to values in the world state for strong data privacy.

Fabric allows for extra-private transactions using a similar conceptual model of off-chain private database as Quorum and Besu for each data collection. As each data collection's peers identity is the same in data collection as on the Fabric network, data collection transaction sent to the shared Fabric ledger reveals the identity of some or all the private transaction's participants, and which contract was executed (the code to the contract is not revealed, but a unique identifier is). This allows other peers on the shared ledger to correlate activity frequency data about other peers' private transactions, in a way that is less subtle than for Quorum and Besu.

Corda's architecture is the least like the other three. The ledger transaction is not a chain of blocks but a chronologically ever-appending ordered list of transactions. Additionally, transactions are only visible in the ledger by the participants, meaning that businesses using different applications are not able to see any statistical transaction information from other applications they are not part of.

Due to the UTXO model used by Corda transactions, participants of a transaction can view previous transactions which are relevant for verifying the current transaction's inputs. This reveals the direct data of that transaction despite not being a participant.

### 5.1.2 Zero knowledge proofs

Nightfall was reviewed to gain insights into the mechanisms behind zero knowledge proofs and their data privacy properties. Users of the nightfall private tokens can hide the recipient and number of transfers; however, the identity of the sender and the frequency of private transactions is obfuscated and may be possible to infer unless another data private solution is paired with private transfers.

### 5.1.3 Private EOSIO chains

Using the EOSIO software to run a private blockchain allows organisations the ability to have strong data integrity guarantees and redundancy while also having full data privacy. Characteristics of the EOSIO are software make solutions extremely flexible, upgradeable, scalable and have many tools and support for application developers.

All peers running a private EOSIO chain gain full visibility of all data on the chain. If different data privacy relationships exist within a system, it is handled by running multiple private EOSIO chains for each data privacy group. EOSIO software does not have built-in support to run multiple EOSIO networks. This may work for a limited number of private data relationships, but complex systems involving many data privacy relationships are not currently possible in any scalable way.

### 5.1.4 Other solutions

Many solutions were reviewed within the EOSIO ecosystem providing different kinds of data privacy solutions. Most of these were proof of concepts. A summary of the problems solved by the solutions:

- Private data communication - sending a private message from one EOSIO account to another.
- Private data storage - the ability to store information split across multiple nodes such that no node can see the data, and that the data can be recovered if all nodes agree. This is intended to be used for distributed private key recovery.
- Sovereign execution environment - a state transition machine that can be used to prove authenticity and traceability of private browser sovereign data.
- Private token transfers - private tokens on the EOS public blockchain

These solutions used hashing, RSA and elliptic curve asymmetric encryption, Shamir secret sharing and other cryptographic primitives to provide data privacy properties and guarantees in conjunction with the use of an EOSIO are blockchain for distributed consensus. They provide a strong reference for the capability and flexibility of using cryptography software in collaboration with a scalable, flexible and programmable blockchain layer to create novel data privacy solutions.

## 5.2 Data privacy strategies

Private transaction managers can manage complex data privacy relationships and create a robust general purpose technology for data privacy with the benefits of a blockchain. By utilizing mature and efficient hashing algorithms, private transaction managers can scale significantly. The need for off-chain multi-party private transaction consensus mechanisms reduces the scalability of private transaction managers compared to the blockchain ledgers that they run on.

No private transaction managers were able to achieve complete data privacy without a data leak in which data is revealed or metadata can be used to analyze trends and statistics about private transactions by unintended viewers. Quorum and Besu had the most subtle and tightest identity metadata leak about private transaction frequency. Corda does not leak metadata, but under certain circumstances leaks direct transaction data. Fabric ledgers are much more private for regular private transactions than Quorum or Besu ledgers but reveal the less subtle identity metadata about extra-private data transactions to peers. Fabric provides multiple mechanisms for data privacy and is the most flexible and complex system of governance and data access controls.

Zero knowledge proofs are a novel new way to solve data privacy issues. They can be theoretically made to be general purpose but are limited due to the significant complexity of setting up zero knowledge engines. The construction and verification of zero knowledge proofs is still significantly less efficient and mature than hashing algorithms. This makes zero knowledge proof solutions less scalable and requires even more specialized developer knowledge and upkeep with current standards to build.

## 5.3 Shared general strategies

### 5.3.1 Cryptography

The use of efficient hashing functions (most commonly sha256) was used effectively in most solutions:
1. All the base layer blockchain protocols for consensus
2. Anchoring private transactions onto shared blockchain ledgers in all private transaction managers
3. Anchoring Blockbase's private database to the EOS or Telos blockchain
4. Anchoring IPFS encrypted data to the shared EOSIO blockchain in Chappe
5. Anchoring private data to the shared EOSIO blockchain in "EOSIO sovereign contract"

Use of asymmetric encryption was a common strategy. This was implemented in EOS communication, Chappe, Sense Chat, and Funnels. Most of these projects are proof of concepts and it should be noted that asymmetric encryption is significantly more inefficient than symmetric encryption making these implementations less scalable.

Pobox and EOS communication implemented the use of a Diffie-Hellman key exchange taking advantage of public key infrastructure while using efficient AES symmetric encryption with a shared key for data transport through the blockchain. As mentioned in Section 3.3, these mechanisms of encrypted persistent and meaningful data on a public blockchain should not be considered private. [83]

The use of TLS was used by all private transaction managers for data transport. This is the transport protocol underpinning https, which was used for most solutions that connected with a browser or other client.

The use of zero knowledge proofs and other advanced cryptography was not found to be in common use in production enterprise systems. This is likely due to its inefficiency and immaturity making it difficult to develop, maintain and less scalable for production systems at the moment. This technology is likely to become more adopted with increased technical maturity, tooling for developers and integration with systems.

### 5.3.2 Coordination via on-chain contracts

The use of smart contracts to coordinate data privacy groups, and more generally to manage identities and access controls was common.

Fabric, Besu, and Quorum use this technique to ensure global coordination of member list and access control. This is a more recent feature for each of them and is likely a continuing trend.

The use of smart contracts for coordination provides reliable synchronization of permissions on peers, and transparency of governance. The shared information, accessible to all peers on the shared ledger, is a level of data transparency that can also be considered private data. This implicit information can present opportunities for the technically enabled to take advantage of depending on how this is implemented.

### 5.3.3 Transaction and contract code sharing

Corda peers are only able to synchronize ledger entries that they are a participant of. This concept of a ledger that may not be synchronized in full by any one peer means that no transaction meta data is linked to unauthorized peers.

Hyperledger Fabric Contract code is only shared with peers that engage in transactions with the code. This implies that other peers on a shared ledger are not privy to the execution logic of business transactions.

EOSIO, Besu and Quorum have a global ledger of transactions and contract code that is shared between all peers meaning that under certain circumstances, this data can be used to infer some additional metadata about private transactions.

### 5.3.4 Data migrations for data privacy applications

As stated at the beginning of this section, solutions stored private data only with relevant parties and used different types of cryptographic primitives to anchor that data on a blockchain so that its integrity can be verified without revealing the data to non-relevant parties.

Due to the raw data not being in plain unencrypted format in a central location this presents an additional challenge when the data or structure of private data needs to be upgraded. In this scenario each participant holding relevant data needs to migrate data separately. This can be automatically done when they update their software or may require a consent flow in which the party consents to the data upgrade. This is different to traditional client/server architectures using a database storing unencrypted values in which the application operator can automatically upgrade all user's data without requiring users to interactively participate.

### 5.3.5 Multiple blockchain ledgers for data privacy

To manage many private data relationships, an equal number of private blockchains can be deployed, each only sharing data between its authorized peers. This strategy was trailed by several clients of original private transaction manager solution providers [1] but was found not to scale. Instead, the use of off-chain private databases anchored to a shared ledger through hashes was found to be more scalable while providing the ability to manage large numbers of complex private debt relationships.

The ability to orchestrate and manage multiple blockchain ledgers each with their own private databases is only provided by Hyperledger Fabric. Managing blockchain nodes and complex data relationships is still technically challenging and requires specialized expertise. The use of automated multi-chain deployment solutions can assist with this. Fabric also provides a cross chain read/write tool as part of the solution to further manage multi chain deployments and provide the most data privacy flexibility.

The use of EOSIO private chains for data privacy was found to be scalable for a limited number of private data relationships. The properties of the EOSIO solution gives clients scalable built-in features to upgrade contract software, easy governance, upgraded blockchain features, advanced key management and other features that make it very applicable for enterprise use cases.

# Appendix A: Analysis framework results

| Technology | Category | Maturity | Private data relationships | Private data transport | Private data integrity | Data privacy |
|---|---|---|---|---|---|---|
| Hyperledger Fabric | Private Transaction Manager - Data Collections | Mature and production ready with relatively significant adoption by enterprise. Open source. Fabric's origination traces back to 2015 when IBM began experimenting with blockchain tech. | Groups of peers on a shared ledger and subgroups appears within a private data collection | Channel level: sent peer-to-peer via encrypted TLS gossip protocol to all peers on the channel. | Channel level: All channel peers have cryptographic proof that other peers have seen and accepted the order of transactions on the ledger. If a peer is in possession of application's contract code, they can additionally execute and verify the transaction status and the state of the contracts data. | Channel level: Only channel peers can see all transactions and their input hashes and read/write set hashes on ledgers. Peers that participate in transactions for shared contracts can see contract code, raw transaction inputs and read/write sets. Transaction sent from private data collection transactions are visible on the ledger and can be attributed to their creator and data collection and only a transaction hash is visible (not inputs or read/write set). Orderer peers can see all channel transaction inputs but are unable to execute transactions and see statuses and state.<br><br>Data collection level: Only data collection peers can see all historic transactions and private contract states. Peers that participate in transactions for shared contracts can see contract code, raw transaction inputs and read/write sets. |
| Corda | Sharded Ledger with Private UTXO Transactions | Mature and production ready with relatively significant adoption by enterprise. Open source with optional enterprise services offered by premium software. R3 was formed in 2014 and made Corda open source in 2016 | Peers using the same application "CorDapp" | Sent directly via encrypted TLS connection exclusively to peers participating in the transaction. | Participants of transactions can get cryptographic proof that other participants have seen, executed and accepted the order of the transaction. This does not necessarily apply to all transactions within a Corda application, as transaction visibility can be defined more granularly than the application. | Only participants of a transaction can see transaction data including all inputs and outputs. Transaction visibility includes previous transaction outputs which they may not have been a participant in. Non-transaction-participants are not able to see any data on the ledger for those transactions. Peers that receive a private transaction can request and will automatically receive all historic transactions leading up to that transaction even if not a participant.<br><br>Network non-validating notaries can see a ordered list of transaction hatches no data of applications that use them. Network validating notaries can see all transaction data for applications that use them. |
| Hyperledger Besu | Private Transaction Manager - Orion | Mature and production ready with relatively significant adoption by enterprise. Began as a technology developed by PegaSys, but joined the Hyperledger family in 2019 | Privacy groups of peers | Sent directly via encrypted TLS connection between peers in the privacy group. | Peers in a privacy group can get cryptographic proof that other peers in the group have seen, executed and accepted the order of private transactions. | Transaction history and private state in a privacy group is visible only to all members of the privacy. This includes visibility of the mapping between private Orion identity and the shared Ethereum network identity. |
| Quorum | Private Transaction Manager - Tessera | Mature and production ready with relatively significant adoption by enterprise. Released in 2016 by JP Morgan to | Private transaction recipients | Sent directly via encrypted TLS connection Two private recipients of the transaction. | Participants of a private transaction can get cryptographic proof that other participants have seen, executed and accepted the order of the private transaction. | Transactions and private state of a private transaction is visible only to participants of the transaction. This includes visibility of the mapping between private Tessera identity and the shared Quorum network identity. |

| | | address financial use cases using blockchain and now managed by Consensys. | | | | |
|---|---|---|---|---|---|---|
| Nightfall | Zero Knowledge Proofs | Beta release in testing with EY's customers. The cryptographical ideas behind zero knowledge proofs have been researched and published on by a global community of academic and industry developers and researchers. EY's Nightfall Was released in mid-2019. | Any user of the Nightfall smart contract on the Ethereum public network. | Private data is processed into a zk-SNARKs zero knowledge proof which conceals any raw data and sent unencrypted to the Ethereum public network. The whisper network is used to anonymously and privately send commitment data to recipients so that they can unlock private transaction outputs. | The smart contract deployed on the public Ethereum network verifies all zero knowledge proves and guarantees the correctness. This allows all nodes of the Ethereum public network to prove that private token transfers and balances have not broken the expected token rules (such as preventing double spends). | The recipient and amount transferred using private tokens can only be viewed by the owner of the private key referenced in the transaction outputs and the owner of the private key that sent the private transaction. Other users on the Ethereum public network can view which Ethereum accounts make private transfers, is very accounts are making transactions and when, but not to whom and how much. All data (sender, recipient and amount) for the public Nightfall token is visible on the Assyrian public network. |

*Table 10: Results of the analysis framework applied to the reviewed non-EOSIO technologies*

| Technology | Category | Maturity | Private data relationships | Private data transport | Private data integrity | Data privacy |
|---|---|---|---|---|---|---|
| EOSIO private chains | Private EOSIO blockchain | Mature and production ready with relatively small adoption by enterprise. Private blockchain is a fully supported by the open source EOSIO core software first released in 2017 by block one. Software to create data privacy rules needs to be built by the application. | Relationships between the users/clients of the blockchain and the different peers is running the EOSIO nodes. | HTTPS connection to the blockchain API or an extended API. Connections between peers can be made using TLS or a VPN tunnel. | Network peers have cryptographic proof that other peers have seen, executed and accepted the order of transactions on the blockchain. | Data privacy between the users/clients depends on the access software run by the private blockchain. This can be configured flexibly to make data privacy rules based on accounts, account groups, transactions, transaction types and more. All operators of the blockchain see all data. |
| Europechain | | Production ready with relatively small adoption by enterprise. Not open source. | | | | Users of Eureopechain can see all blockchain data except for any data associate with "forgotten" accounts. All operators of the blockchain see all data. |
| Blockbase | Pegged private blockchain database | Beta release in 2020 with active research done since 2018. | Relationships between the users/clients of the blockchain. | | | |

| Name | Category | Status | Private data participants | Private data transport | Private data integrity | Privacy properties |
|---|---|---|---|---|---|---|
| Pobox EOS communication | Private data communication: encrypt data | Proof of concept. Open source | Between sender and receiver EOSIO accounts. | HTTPS connection to the network API. Data is encrypted using a Diffie Hellman key exchange for transit over the blockchain. | Participants of data communication have cryptographic proof of the contents ordering and author of the data. | Only participants in the communication see message content. The encrypted message content is forever on the blockchain and may be decrypted if quantum computing matures. Other peers on the network can see the time of message, sender and recipient but is not the message content. |
| Chappe | Private data communication: encrypt data hash | Proof of concept. Open source | Between sender and receiver EOSIO accounts. | Encrypted data message is sent via IPFS with a hash sent via the blockchain. | Participants of data communication have cryptographic proof of the contents ordering and author of the data. | Only participants in the communication see message content. The encrypted message content is forever on the blockchain and may be decrypted if quantum computing matures. Other peers on the network can see the time of message, sender and recipient but is not the message content. |
| Sense Chat | Private data communication: encrypt communication handshake | In production use on EOS blockchain since 2018. Not open source. | Relationship between sender and receiver of private data. | webRTC connection, brokered on the blockchain. | Participants of data communication have cryptographic proof of the data connection handshake. | Only participants in the communication see message content. Other peers on the network can see the session starts time, sender and receiver. This does not include message timestamps but only the start of each webRTC session. |
| Funnels | | Proof of concept. Open source. | Relationship between sender and receiver of private data. | webRTC connection, brokered on the blockchain. | | |
| PrivEOS | Private data storage | Ready for operation but not in use. Open source. | Between users and the operators. | HTTPS connection to the network API. | Users can get cryptographic proof that a private key is stored by nodes on the network. | The stored private key is only visible in small parts to each node in the network. It appears in the networks collaborate they can put all the pieces together to form the private key. |
| EOSIO sovereign contract | Private high integrity data | Proof of concept. Open source | Between different user clients that wish to share sovereign data. | All | If sovereign data is shared with another user, they can get cryptographic proof of the data content and its history. | Only the sovereign contract user can see the private data. Users who are explicitly shared this data are to be also able to view it. |
| EOSVM | | EOSVM is part of the open source EOSIO core software, As of January 2020 and in use on several production blockchains. | | | | |
| EOS Blender | 2nd Layer | In production use on EOS blockchain since 2019 with minimal usage. Not open source. | Sender and receiver of EOS tokens. | HTTPS connection to the network API. | All deposit and withdrawal transactions are transparent and on the public EOS blockchain with proof of transaction and order. Private data does not have any data integrity properties that can be used by the users. | The EOS Blender operators retain all private data about transfers and provide the cryptographic hashes necessary to generate transactions. |
| Peos | | Alpha phase and waiting for code changes in the EOSIO to be able to launch on production chains. | Sender and receiver of pEOS tokens. | HTTPS connection to the network API. | Participants of a pEOS private token transfer have cryptographic proof of the transaction sender, recipient and amount and the order of transactions. | Sender, recipient and amount transferred is private between private participants of private transfers. The identity and activity of the EOS account that makes the transfer requests are public. |

*Table 11: Results of the analysis framework applied to the reviewed EOSIO technology*